\documentclass[twocolumn]{aastex631}
\usepackage{amsmath}

\newcommand\teff{\ensuremath{T_{\rm eff}}}

\newcommand\Lx{\ensuremath{L_X}}

\newcommand{\lbol}{\ensuremath{L_{\rm bol}}}
\newcommand{\prot}{\ensuremath{P_{\rm rot}}}

\defcitealias{kounkel2018a}{Paper I}

\begin{document}
\title{The evolution of stellar X-ray activity and angular momentum as seen by eROSITA, TESS, and Gaia}

\author[0000-0002-3481-9052]{Keivan G.\ Stassun}
\affiliation{Department of Physics and Astronomy, Vanderbilt University, Nashville, TN 37235, USA}
\author[0000-0002-5365-1267]{Marina Kounkel}
\affiliation{Department of Physics and Astronomy, University of North Florida, 1 UNF Dr, Jacksonville, FL, 32224, USA}

\begin{abstract}
We have assembled a sample of $\sim$8200 stars with spectral types F5V--M5V, all having directly measured X-ray luminosities from eROSITA and rotation periods from TESS, and having empirically estimated ages via their membership in stellar clusters and groups identified in {\it Gaia\/} astrometry (ages 3--500~Myr). This is the largest such study sample yet assembled for the purpose of empirically constraining the evolution of rotationally driven stellar X-ray activity. We observe rotation-age-activity correlations that are qualitatively as expected: stars of a given spectral type spin down with age and they become less X-ray active as they do so. We provide simple functional representations of these empirical relationships that predict X-ray luminosity from basic observables to within 0.3~dex. Interestingly, we find that the rotation-activity relationship is far simpler and more monotonic in form when expressed in terms of stellar angular momentum instead of rotation period. We discuss how this finding may relate to the long-established idea that rotation-activity relationships are mediated by stellar structure (e.g., convective turnover time, surface area). Finally, we provide an empirical relation that predicts stellar angular momentum from basic observables, and without requiring a direct measurement of stellar rotation, to within 0.5~dex. 
\end{abstract}


\section{Introduction}

Low-mass stars exhibit phenomena that reveal the presence of magnetic fields at their surfaces, such as periodic variations in their light curves due to magnetic starspots \citep[e.g.][]{bouvier1989} and X-ray emission due to magnetically heated coronae \citep[e.g.,][]{feigelson1999}. These magnetic fields are generally strongest when the stars are very young and rotating rapidly, as evinced by very strong X-ray emission. The fields steadily decline in strength with age, as magnetized winds carry away stellar angular momentum, resulting in stellar spin-down, weakened magnetic dynamo action, and evinced by declining X-ray activity. 


One consequence of these behaviors is that it is possible, in principle, to estimate the ages of stars by measuring the strength of their X-ray emission \citep[e.g.,][]{getman2023} or by measuring their rotation periods \citep[e.g.,][]{barnes2003}, assuming well-calibrated empirical relationships between stellar rotation, X-ray luminosity, and age, as functions of basic stellar properties such as mass. 

Until recently, the data needed to calibrate these relations have been fairly sparse, limited to only a small number of star clusters of known ages that have been observed by X-ray telescopes with limited field of view, coupled with rotation periods measured from (usually ground-based) high-cadence optical photometry of a small fraction of the stars within these X-ray observed clusters. Now, major advances have been made on all of these fronts, enabling the empirical determination of rotation-activity-age relationships with stellar samples of unprecedented size and quality. 

First, space-based surveys such as TESS are now able to provide high-cadence, ultra-precise photometry of stars across most of the sky, enabling the measurement of stellar rotation periods by the tens of thousands \citep[e.g.,][]{kounkel2022a}. 
Second, the eROSITA X-ray observatory has now made publicly available its first data release \citep[DR1;][]{merloni2024,freund2024}, which provides precise X-ray fluxes for stars across half of the entire sky. 
Finally, and importantly, where the determination of stellar ages has traditionally been limited to those that are known to be members of well-established and well-studied clusters, the advent of {\it Gaia\/} and its ultra-precise astrometry (distances and kinematics) for over a billion stars across the entire sky has made it possible to identify thousands of new clusters, moving groups, and even very widely distributed associations \citep[e.g.,][]{kounkel2020}, significantly increasing the census of stars with known ages. 

This paper represents a first effort to assemble the largest sample of stars possessing all of the above data, with the goal of empirically constraining the evolution of stellar X-ray activity and angular momentum, and to empirically calibrate widely used rotation-activity-age relationships. Section~\ref{sec:data} presents the data used in this study and the methods used to derive key quantities from them. Section~\ref{sec:results} presents the key results of this investigation, and some implications of these findings are discussed in Section~\ref{sec:disc}, which also presents a summary of the main conclusions.

\section{Data and Methods}\label{sec:data}

\subsection{Sample Selection for this Study}\label{subsec:sample}
We sought to assemble the largest possible sample of stars possessing all three of the key measurable quantities required for this investigation, and for which all three quantities were determined from entirely independent data sources. In particular, we selected stars from the union of: 
\begin{enumerate}
    \item X-ray fluxes newly reported in the eROSITA first data release \citep[DR1;][]{Merloni:2024}, which when combined with parallaxes from {\it Gaia\/} DR3 yields the X-ray luminosity, \Lx; 
    \item Periodic light curve variations identified in the TESS full-frame image (FFI) data by \citet{kounkel2022b}, which we take to be the stellar rotation period unless otherwise indicated; and 
    \item Ages estimated from the stars' membership in stellar clusters, groups, and associations within 3~kpc identified in {\it Gaia\/} by \citet{kounkel2020}. 
\end{enumerate}
The result is a set of 8267 stars with \teff\ in the range 3000--6700~K (corresponding approximately to spectral types F5--M5), periodic signals in the range 0.1--20~d (set by the cadence and duration of typical TESS light curves), \Lx\ in the range $10^{27.8}$--$10^{31.5}$ erg~s$^{-1}$ (set by the combination of eROSITA sensitivity and the underlying emission mechanisms), and ages in the range 3--500~Myr \citep[set by the range of applicability of the cluster/group identification method developed by][]{kounkel2020}. This is the largest such sample of stars with these properties heretofore assembled. 

\subsection{Calculating Derived Stellar Properties}

In addition to the directly measured or inferred quantities from Section~\ref{subsec:sample}, we also require estimates of stellar effective temperature (\teff), as well as radius ($R$) and mass ($M$) from which, together with the rotation period (\prot), we can calculate angular momentum ($L$). 

We adopt \teff, $M$, and $R$ of the stars from the photometric estimates in the TESS Input Catalog \citep{stassun2019}, as we did for this same study sample in our previous work \citep{kounkel2022b}. 

Finally, for $L$, we treat the stars as simple solid body rotators, for which $L = 2/5 M R^2 \omega$, where $\omega$ is the angular velocity transformed from \prot. This is not a perfect approximation, and there are likely \teff-dependent variations to the formalism of the stellar moment of inertia. However, utilizing more advanced stellar interiors models for the moment of inertia \citep[e.g.,][]{baraffe2015} does not have a qualitative impact on our observations, and in fact increases the scatter in the observed relations over the simple approximation that we have adopted.

\subsection{Catalog Availability}

The catalog resulting from the combination of directly measured and derived stellar properties for the entire stellar sample is available via the Filtergraph data portal at \url{https://filtergraph.com/erosita_tess_gaia_v1}.

\section{Results}\label{sec:results}

\begin{figure}[!t]
\includegraphics[width=\linewidth]{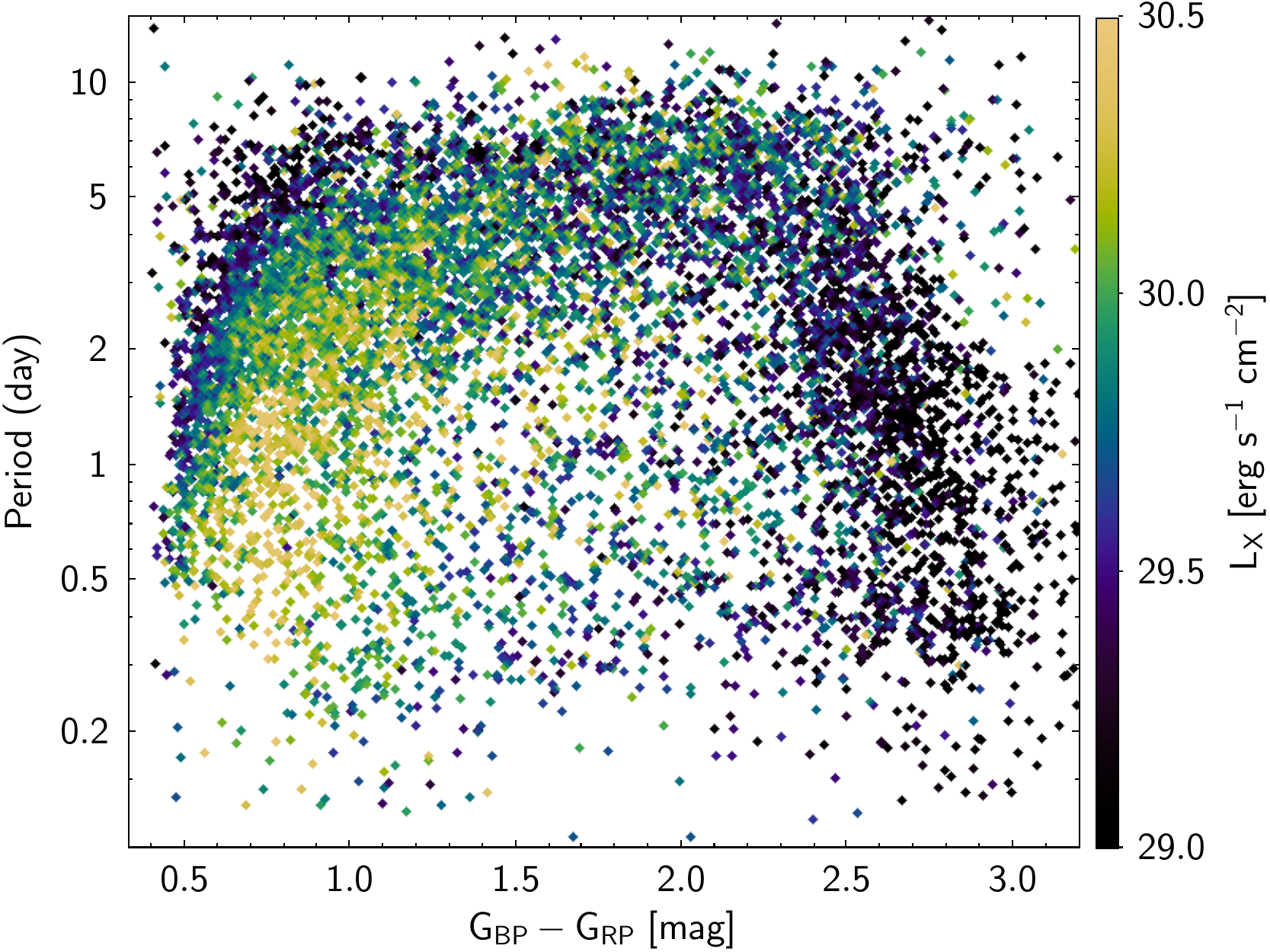}
\includegraphics[width=\linewidth]{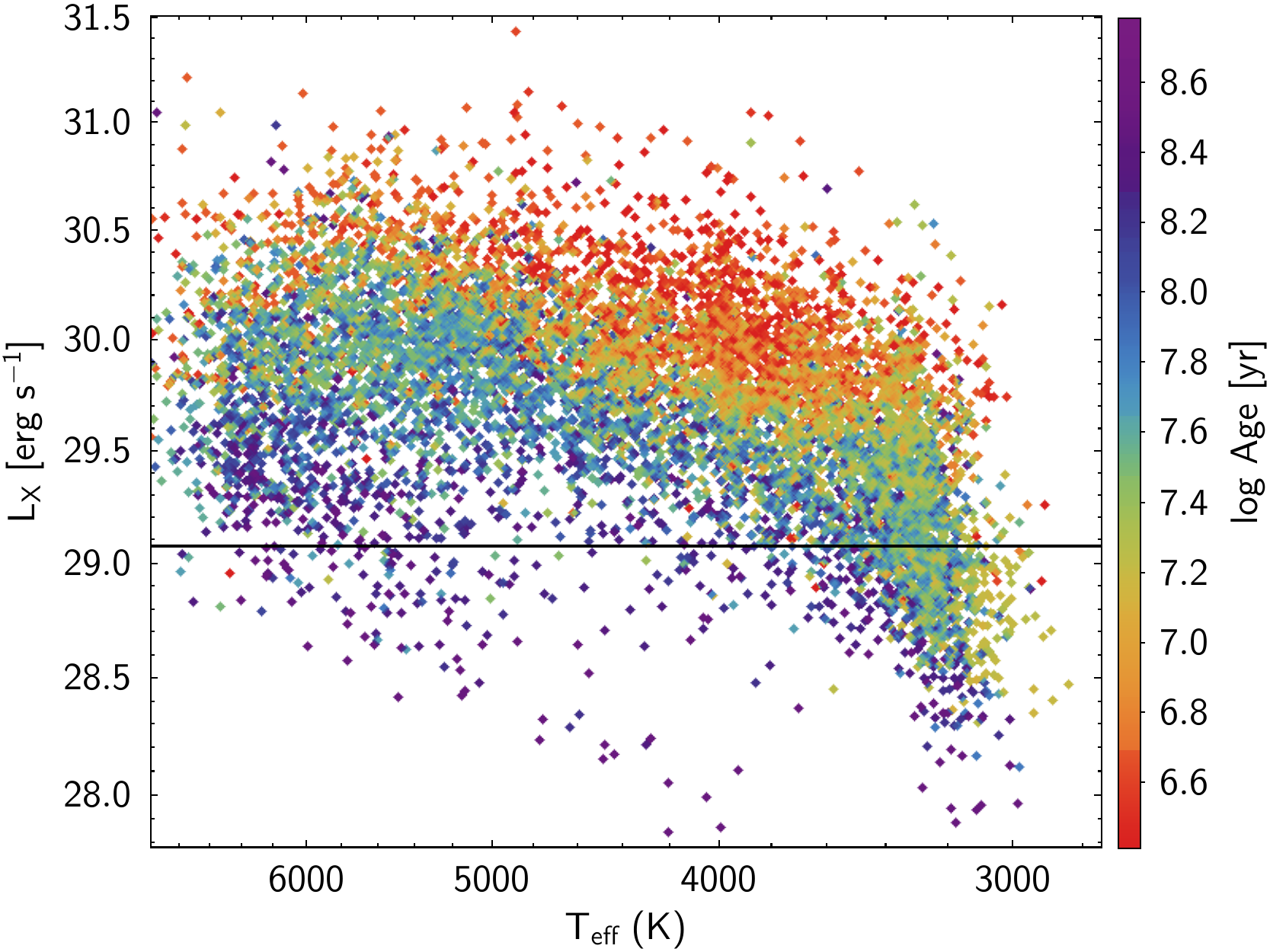}
\caption{(Top:) Gyrochronology diagrams of the study sample represented by \prot\ versus stellar color.
\Lx, which correlates with stellar age, is represented by color. The usual ``horseshoe shape" of relatively old (X-ray faint) and relatively slowly rotating stars defines the upper envelope (darker symbols), with relatively younger stars (X-ray bright) ranging down into the shorter rotation periods (lighter symbols). 
(Bottom:) Relationship of \Lx\ with stellar ages determined independently by \cite{Kounkel:2023}. As expected, the younger stars at a given \teff\ generally exhibit stronger \Lx. {The eROSITA flux sensitivity limit at 200~pc, corresponding to the typical distance of stars in our study sample, is shown as a black line.}
\label{fig:Prot}}
\end{figure}

One important application of stellar rotation period measurements is for calibration of empirical gyrochronology relations that link stellar rotation and activity to age \citep[e.g.,][]{mamajek2008}. Figure~\ref{fig:Prot} (top) represents the \prot\ of the study sample as a function of stellar color, in which we observe the familiar ``horseshoe" morphology that has been reported in numerous works. The common interpretation is one in which stars of a given color or \teff\ (i.e., mass) evolve upwards in the diagram toward longer \prot, as they steadily lose angular momentum with time due to the effects of magnetized stellar winds \citep[see, e.g.,][]{barnes2007}. 


The fact that the stars along the upper envelope of Figure~\ref{fig:Prot} are the least X-ray luminous is consistent with the expectation that stars become less magnetically active as they spin down with age. Figure~\ref{fig:Prot} (bottom) shows this relationship more directly, where \Lx\ as a function of \teff\ clearly evolves with stellar age in the expected sense. 
Here, we have removed stars having $P_{\rm rot} < 2$~d, as \citet{kounkel2022b} demonstrated that the large majority of such stars in this sample are likely to be binaries on the basis of their position on the ``binary main sequence" in the Hertzsprung-Russell diagram and on the basis of having apparently oversized radii (due to the light contribution of a companion star); a similar interpretation for stars with $\prot < 2$~d has also been reported by other authors \citep[e.g.,][]{douglas2016,douglas2017,stauffer2018,simonian2019}. 

{Figure~\ref{fig:Prot} (bottom) also shows the stated flux sensitivity limit for eROSITA, translated to \Lx\ for the distance (200~pc) out to which the stars in our sample are representative of the underlying population, for the full range of ages and masses spanned by our sample (see Section~\ref{subsec:sample}); at distances greater than 200~pc, our sample is dominated by the Orion Nebula population (typical ages $\lesssim$10~Myr) at $\sim$400~pc. As can be seen from Figure~\ref{fig:Prot}, the eROSITA sensitivity limit results in good completeness to 200~pc, except for stars cooler than $\sim$3400~K at ages older than $\sim$100~Myr. Because of the dominance of the Orion Nebula population at larger distances, the sample is overrepresented by stars with ages 3--10~Myr, well above the \Lx\ sensitivity limit.}

\begin{figure*}[!t]
    \centering
    \includegraphics[width=\linewidth]{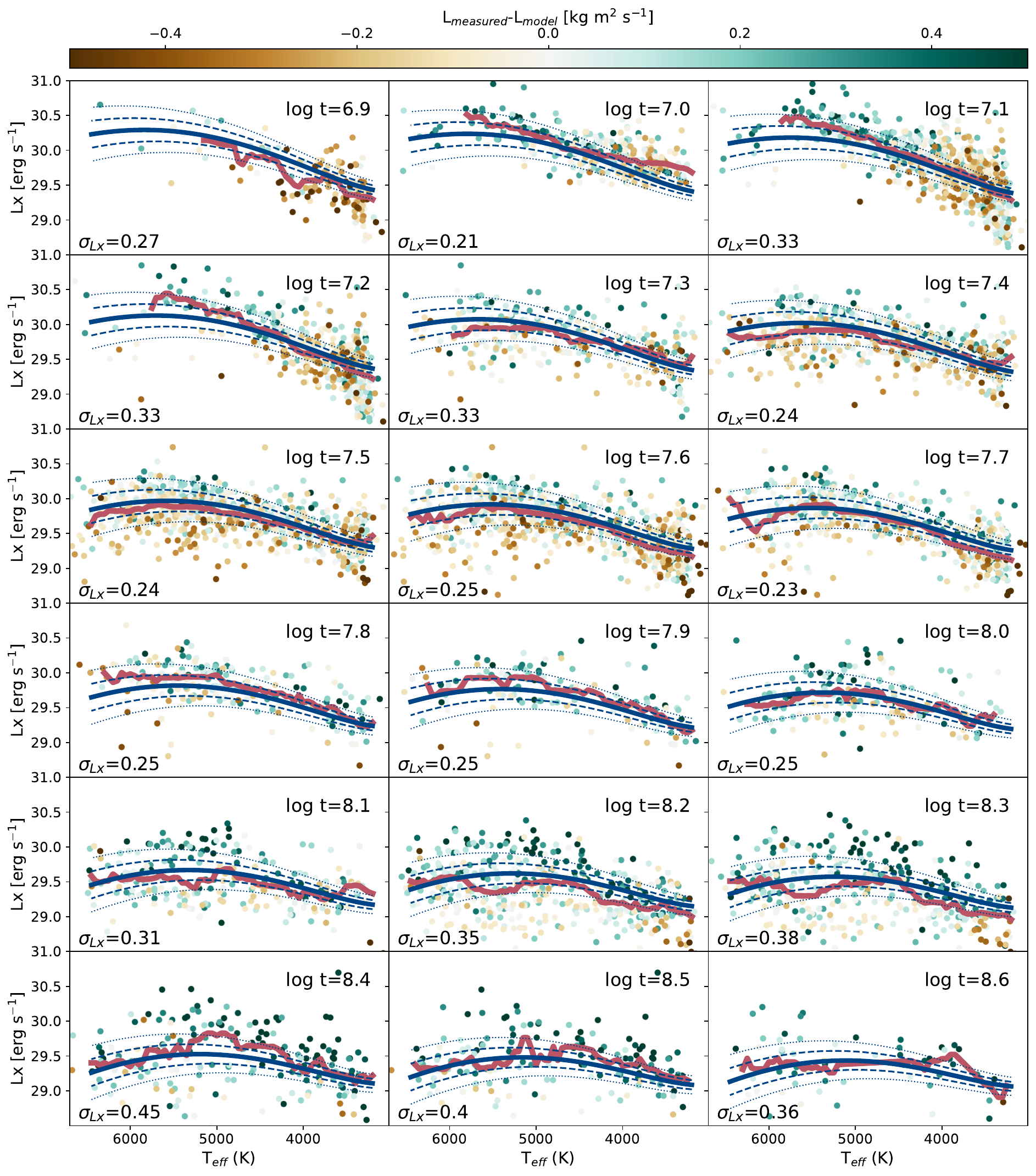}
    \caption{Relationship of \Lx\ as a function of \teff\ across age for stars in 0.1~dex age bins. The red line corresponds to the running median of $L_x$ as a function of \teff\ at a given age. The best fitted-model for a given age is shown in the solid blue line. Other lines show the trend of the model with respect to age: in each panel, dashed line is the model with $\pm0.3$ dex difference from the given age, and the dotted line is $\pm0.6$ dex difference. The scatter between the data and the model for a given age bin is shown in the corner. The symbols represent individual stars in the sample and are colored according to their angular momenta ($L$) relative to the median $L$ in each age bin. It is evident that $L$ contributes significantly to the apparent scatter in \Lx.}
    \label{fig:fit_v_teff}
\end{figure*}

The relationships observed between \Lx, \teff, and age can be represented in functional form. In Table~\ref{tab:coeff} we report the best-fit coefficients for $\Lx = f(\teff,t)$ according to Equation~\ref{eqn1}: 
\begin{equation}\label{eqn1}
\begin{split}
\log L_X=a_0+a_1\log T_{\rm eff}+a_2(\log T_{\rm eff})^2+a_3(\log T_{\rm eff})^3+\\
b_0\log t +b_1\log t\log T_{\rm eff}+b_2\log t(\log T_{\rm eff})^2
\end{split}
\end{equation}
The typical scatter in \Lx\ around this empirical relation is $\sim$0.3~dex across the age range spanned by the study sample (see Figure~\ref{fig:fit_v_teff}).

\begin{figure}[!t]
\includegraphics[width=\linewidth]{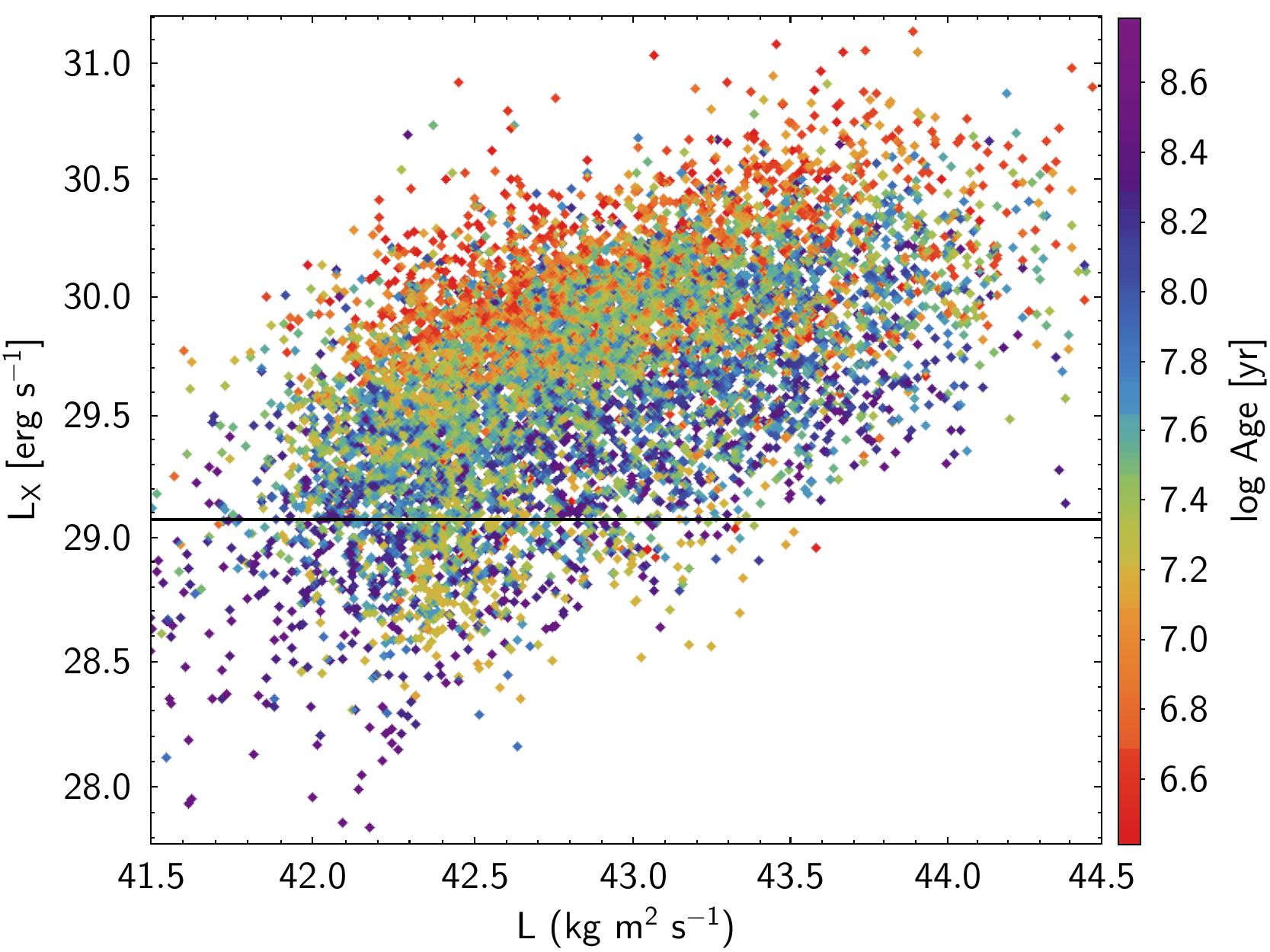}
\includegraphics[width=\linewidth]{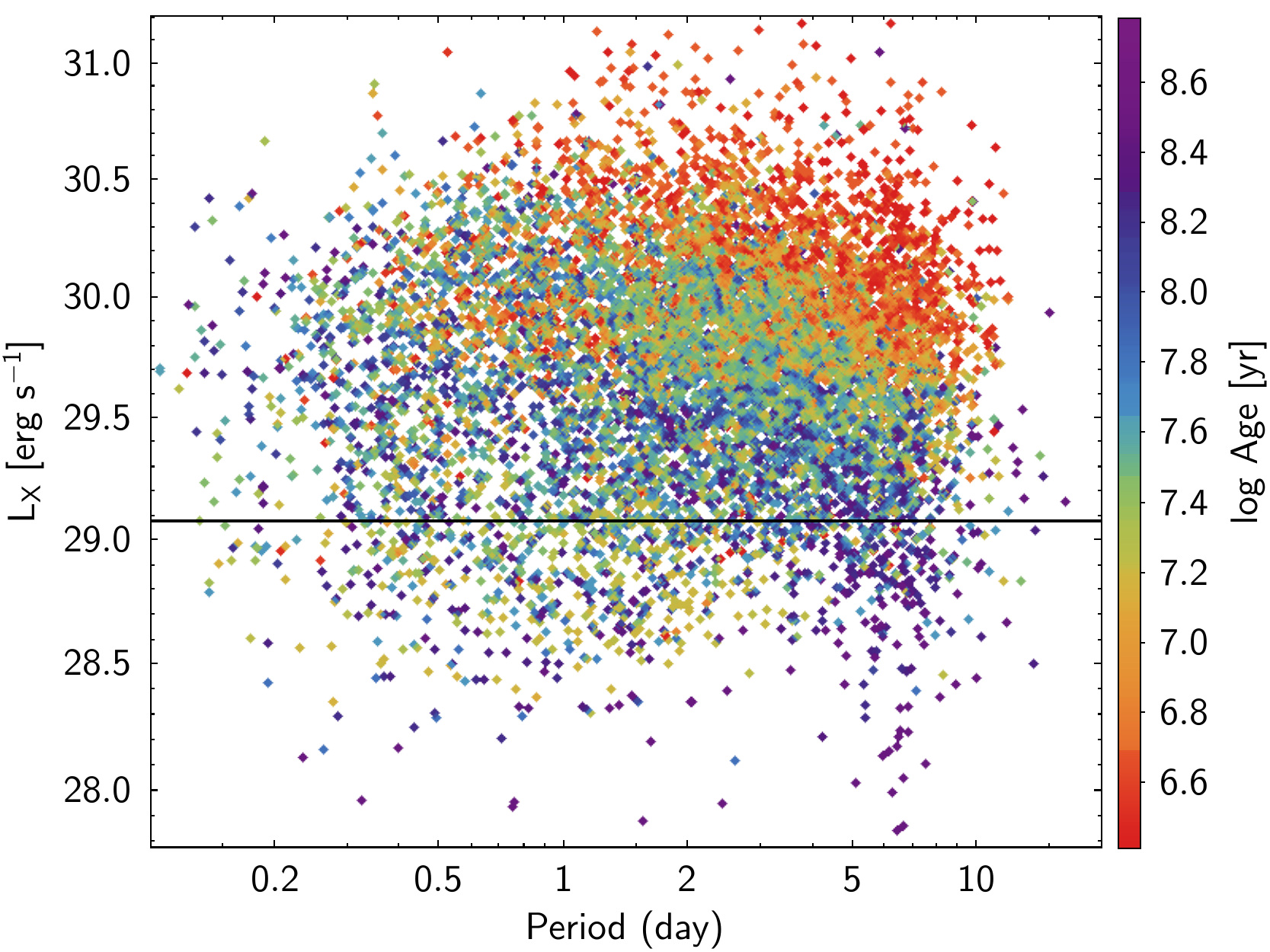}
\caption{(Top) Relationship of \Lx\ with stellar angular momentum, $L$. As in Figure~\ref{fig:Prot}, \Lx\ varies with stellar age. However, it is also clearly apparent that, for stars of a given age, \Lx\ correlates strongly---and in a very simple manner---with $L$, and this \Lx--$L$ relationship is much simpler and tighter than the \Lx--\prot\ relationship (bottom). {The horizontal line in each panel is the same as in Figure~\ref{fig:Prot}.}
\label{fig:Lx_angmom_age}}
\end{figure}

Next, with the knowledge that stellar rotation is a key driver of X-ray activity, Figure~\ref{fig:Lx_angmom_age} represents \Lx\ as the functional dependent on \prot\ (bottom panel) or $L$ (top panel), and stars of a given age represented by the color gradient. While the gyrochronological effect (age evolution of \Lx) is evident in the case of \prot\ as the independent variable, it is notable and striking how much more evident it is in the case of $L$ as the independent variable, the functional dependence being clearly simpler and more monotonic. Indeed, in Figure~\ref{fig:fit_v_teff} it was already evident that $L$ is a significant contributor to the apparent scatter in the \Lx\ versus \teff\ relationship, confirming that $L$ must be an important variable.

\begin{figure}[!t]
    \centering
    \includegraphics[width=\linewidth]{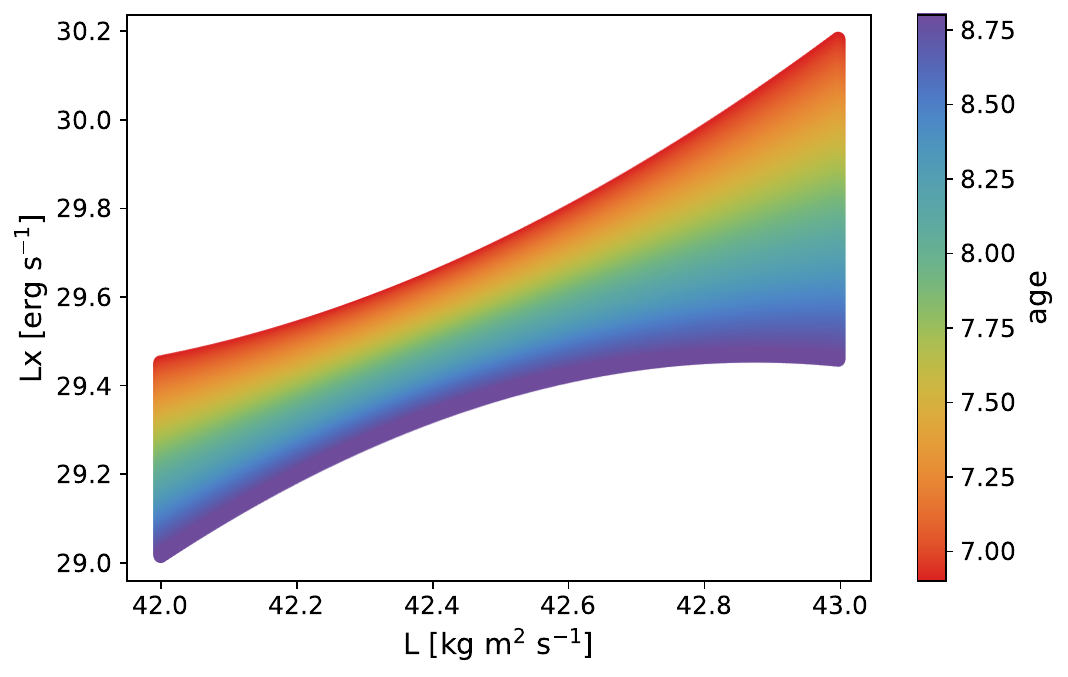}
    \includegraphics[width=\linewidth]{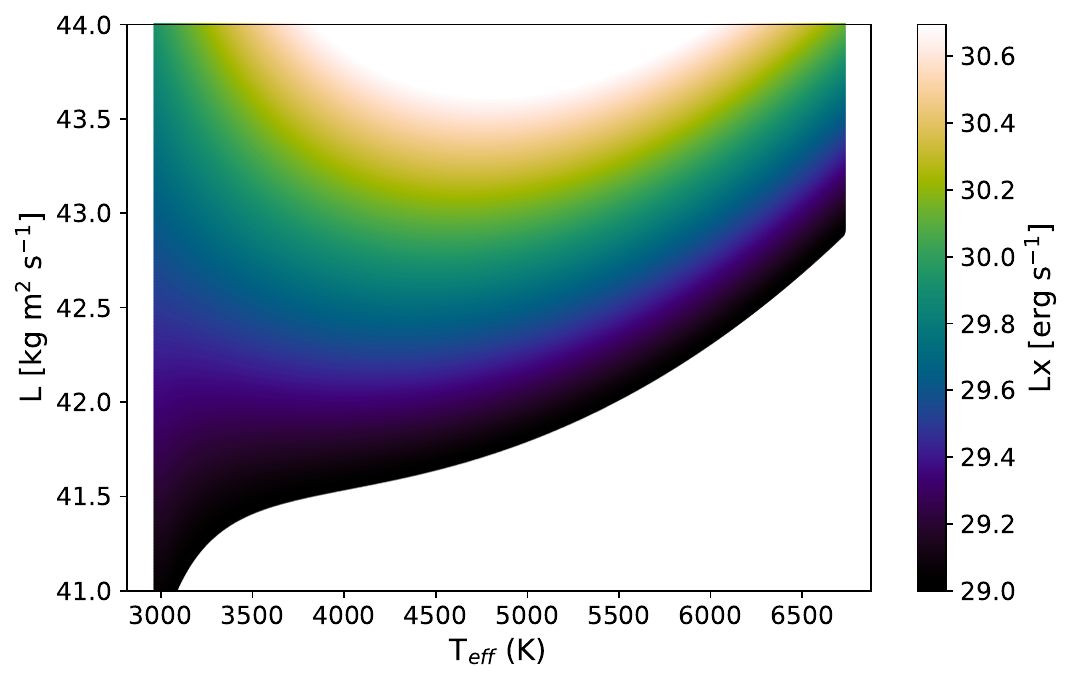}
    \caption{(Top:) Representation of the full model fit (Equation~\ref{eqn2}) for \Lx\ across the full range of $L$ and stellar age. (Bottom:) Same as top, but for $L = f(\Lx, \teff)$ according to Equation~\ref{eqn3}.}
    \label{fig:L_model}
\end{figure}

Given the apparent functional simplicity of \Lx\ versus $L$, we performed another fit akin to that in Equation~\ref{eqn1}, but now representing the expected \Lx\ as a function of age and $L$: 
\begin{equation}\label{eqn2}
\begin{split}
\log L_X=a_0+a_1\log L+a_2(\log L)^2+a_3(\log L)^3+\\
b_0\log t +b_1\log t\log L+b_2\log t(\log L)^2
\end{split}
\end{equation}
For simplicity of visualization, the full model for \Lx\ across the full range of $L$ and age is represented in Figure~\ref{fig:L_model}. 


Finally, the interrelationships demonstrated above between \Lx, \teff, and $L$ suggest that it is now possible to estimate a star's $L$ with knowledge of just two observables that are now available for a large number of stars, regardless of whether they are known to be members of clusters or associations with known ages. Similarly to the above we can fit $\Lx = f(\teff, L)$, and, since the expression is linear with respect to $L$, the expression is thus reversible, and, even with no estimate of a star's age, we may express $L = f(\Lx, \teff)$, according to the following relationship: 
\begin{equation}\label{eqn3}
\begin{split}
\log L=(\log\Lx - a_0 -a_1\teff -a_2\teff^2 -a_3\teff^3)/ \\
(b_0 +b_1\teff +b_2\teff^2) 
\end{split}
\end{equation}
where the fitted coefficients are again provided in Table~\ref{tab:coeff}. 
The typical scatter of $L$ about the fitted relationship is $\sim$0.5~dex.

{All three parameters ($L$, $\teff$, and age) play a very important role in predicting $\Lx$. Including all of these parameters in a single fit becomes prohibitively complex for a simple polynomial, but this highlights that all three fits are highly complementary to one another, exploring different aspects of the parameter space. In Table~\ref{tab:coeff} we calculate the reduced $\chi^2$ ($\chi_\nu^2$) for each interpolation, having a value of $\approx$0.11 in each case, indicating that the goodness of fit of all three representations are comparable.}

For visualization purposes, the smooth functional relationship is depicted in Figure~\ref{fig:L_model} (bottom). 
We observe that the overall range of $L$ is largest among the coolest stars, and is smallest among the hotter stars. This is consistent with the general finding from gyrochronology studies that the hotter stars coalesce more quickly onto a ``slow rotator sequence" \citep[sometimes referred to as the ``I sequence"; see, e.g.,][]{barnes2007}. In contrast, gyrochronology studies of low-mass M-dwarfs have observed that the timescale for the development of a clear slow-rotator sequence is much longer \citep[e.g.,][]{douglas2019}. These trends were implied already in the \prot\ data (Figure~\ref{fig:Prot}), but the data now available here allow these trends to be more clearly understood in the context of angular momentum itself.

\begin{deluxetable}{cccc}
\tablecaption{Fitted coefficients for \Lx\ as a function of different combinations of age, \teff\ or $L$ \label{tab:coeff}}
\tabletypesize{\footnotesize}
\tablewidth{\linewidth}
\tablehead{
 \colhead{Coefficient} &
 \colhead{\teff\ \& age (Eq.~\ref{eqn1})} &
 \colhead{$L$ \& age (Eq.~\ref{eqn2})} &
 \colhead{\teff\ \& $L$ (Eq.~\ref{eqn3})}
}
\startdata
$a_0$ & 3429.2006 & 9664.4179 & 6904.8952  \\
$a_1$ & $-$2781.3971 & $-$510.6559 & $-$4640.3441  \\
$a_2$ & 755.4020 & 7.9881 & 1010.7142  \\
$a_3$ & $-$68.0371 & $-$0.030732 & $-$70.7200 \\
$b_0$ & $-$28.2623 & $-$948.7022 & $-$87.5350  \\
$b_1$ & 16.6589 & 44.7907 & 46.3483  \\
$b_2$ & $-$2.4700 & $-$0.52876 & $-$6.0646  \\
\hline\hline
$\chi_\nu^2$ & {0.105} & {0.119} & {0.109}
\enddata
\end{deluxetable}

\section{Discussion and Conclusions}\label{sec:disc}

Our finding that stellar X-ray luminosity correlates with angular momentum in a manner that is simpler and more monotonic than with rotation period (Figure~\ref{fig:Lx_angmom_age}) was unexpected. The long-standing explanation for the observed relationship between stellar activity and stellar rotation is that the activity reflects the strength of the stellar magnetic field, and most dynamo mechanisms for field generation depend most directly on the stellar rotation period, not a star's total angular momentum content, which involves mass and radius.

There have been suggestions in the literature that the stellar surface area may be relevant to a star's overall \Lx\ if, for example, the total X-ray production scales with the area available for X-ray emitting magnetic structures \citep[see, e.g.,][]{Stelzer:2016,Jeffries:2011}. 
{Such a surface-area effect could appear as an especially strong correlation with \Lx\ if the sample were entirely in the ``saturated" regime, with roughly constant \Lx/\lbol. We checked the distribution of \Lx/\lbol\ for our sample, finding as expected that the stars of a given mass evolve within the age range of our sample (3--500~Myr; see Section~\ref{sec:data}) from a roughly fully saturated state to a non-fully saturated state \citep[see also, e.g.,][]{getman2023}.}
Therefore, we performed an alternate model fit to Equation~\ref{eqn2} with surface area in place of $L$, finding that the resulting scatter of \Lx\ about the model is approximately the same. Of course, both $L$ and surface area involve the same factor of $R^2$, so this similarity in model fit does not by itself help to disambiguate between them. 

On the other hand, it is expected that properties of a star's interior structure should factor into and mediate the relationship between rotation and activity. In particular, in the standard $\alpha$-$\Omega$ dynamo mechanism, convective action in the stellar envelope (often parametrized by the convective overturn timescale, $\tau_C$) provides the electric ``current loop" that, together with the star's rotation, is responsible for the generation of the star's magnetic field. Consequently, empirical rotation-activity relationships are often found to be tightest when the rotation parameter incorporates $\tau_C$, such as with the Rossby number, $R_0 \equiv \tau_C / \prot$. 
And because $\tau_C$ depends on global stellar properties such as mass (and by extension to global properties that correlate with mass), it may not be surprising that the rotation-activity relationship will also become tighter when the rotation parameter incorporates those global properties, as is the case with $L$. 

The observation that $L$ represents a simpler descriptor than \prot\ for understanding the evolution of stellar rotation was noted by \citet{kounkel2022b} in the context of empirical gyrochronology relations. The present work extends that observation to empirical rotation-activity relations as well. 

{Regardless of their underlying physical cause,} the empirical relations presented here provide the means for predicting stellar \Lx\ on the basis of $L$ and age, and conversely the means to predict $L$ from direct observables such as \teff\ and \Lx, without requiring knowledge of the star's rotation. Moreover, the empirical and quantitative measurement of $L$ and \Lx\ presented here for a very large sample of stars across a large range of masses and ages should serve as observational touchstones for theoretical models of the evolution of stellar X-ray activity and angular momentum.



\section*{Acknowledgments}
This work has made use of data from the European Space Agency (ESA)
mission {\it Gaia} (\url{https://www.cosmos.esa.int/gaia}), processed by
the {\it Gaia} Data Processing and Analysis Consortium (DPAC,
\url{https://www.cosmos.esa.int/web/gaia/dpac/consortium}). Funding
for the DPAC has been provided by national institutions, in particular
the institutions participating in the {\it Gaia} Multilateral Agreement.

\software{TOPCAT \citep{topcat}}

\bibliography{references.bib,main_alt.bib,kgs.bib}

\end{document}